\documentclass[aps,prl,a4paper,reprint,nofootinbib,superscriptaddress,preprintnumbers]{revtex4-1}

\usepackage{hyperref}
\usepackage{amsmath}
\usepackage{amsfonts}
\usepackage{amssymb}
\usepackage{color}
\usepackage[utf8]{inputenc}
\usepackage{graphicx}
\usepackage{microtype}
\usepackage{siunitx}
\usepackage{todonotes}

\newcommand{\slcomment}[1]{
\ifnotwordcount
{\large \color{blue} SL: #1}
\fi}

\newif\ifnotwordcount
\notwordcounttrue

\newcommand{\sw}[1]{\textsc{#1}}
\newcommand{\header}[1]{\textbf{#1:}}
\newcommand{\theoryparam}[0]{\boldsymbol{\theta}}

\begin{document}
\preprint{CTPU-16-35}
\preprint{IFT-UAM/CSIC-16-116}
\title{Accelerating the BSM interpretation of LHC data with machine learning}

\date{\today}

\author{Gianfranco Bertone}
\affiliation{GRAPPA, University of Amsterdam, Science Park 904, 1098 XH Amsterdam, Netherlands}
\author{Marc Peter Deisenroth}
\affiliation{Department of Computing, Imperial College London, 180 Queen's Gate, SW7 2AZ London, United Kingdom}
\author{Jong Soo Kim}
\affiliation{Center for Theoretical Physics of the Universe, Institute for Basic Science (IBS), Daejeon, 34051, Korea and \\
Instituto de F\'isica Te\'orica UAM/CSIC, Madrid, Spain}
\author{Sebastian Liem}
\affiliation{GRAPPA, University of Amsterdam, Science Park 904, 1098 XH Amsterdam, Netherlands}
\author{Roberto Ruiz de Austri}
\affiliation{Instituto de F\'isica Corpuscular IFIC-UV/CSIC, Valencia, Spain}
\author{Max Welling}
\affiliation{Informatics Institute, University of Amsterdam, Science Park 904, 1098 XH Amsterdam, Netherlands}

\begin{abstract}
The interpretation of Large Hadron Collider (LHC) data in the framework of Beyond the Standard Model (BSM) theories is hampered by the need to run computationally expensive event generators and detector simulators.
Performing statistically convergent scans of high-dimensional BSM theories is consequently challenging, and in practice unfeasible for very high-dimensional BSM theories.
We present here a new machine learning method that accelerates the interpretation of LHC data, by learning the relationship between BSM theory parameters and data. As a proof-of-concept, we demonstrate that this technique accurately predicts natural SUSY signal events in two signal regions at the High Luminosity LHC, up to four orders of magnitude faster than standard techniques. The new approach makes it possible to rapidly and accurately reconstruct the theory parameters of complex BSM theories, should an excess in the data be discovered at the LHC.

\end{abstract}

\ifnotwordcount
\maketitle
\fi

\header{Introduction} 
A vast effort is currently in progress to discover physics Beyond the Standard Model (BSM) at the Large Hadron Collider (LHC), motivated in part by the possible connection between new particles at the weak scale and the dark matter problem in astrophysics and cosmology~\cite{Jungman:1995df,Bertone:2004pz,Bertone:2010at}.
The absence of clear evidence for BSM physics in current LHC data has been interpreted in the context of simplified models~\cite{Alves:2011wf,Abdallah:2015ter} as well as of {\it full} models, such as various incarnations of the minimal Supersymmetric Standard Model (MSSM)~\cite{Strege:2014ija, Aad:2015baa, deVries:2015hva, Aaboud:2016wna, Khachatryan:2016nvf}. 

Such studies, and even more the interpretation of a hypothetical excess in future data, are hampered by  the computationally intensive task of sampling the high-dimensional parameter space of theoretical models, and comparing, for each sample, the predicted signal with actual data.
For each set of input parameters one needs in fact to: (i) generate a Monte Carlo (MC) sample of collision events; (ii) run the sample through a detector simulation; and (iii) compare the predicted signal with data, often within signal regions (SRs) defined by experimental cuts on observable quantities, such as missing transverse energy, number of jets, momenta, and angles \cite{Aad:2015baa}.  
This procedure is computationally very expensive, and it constitutes the bottleneck for global analyses of BSM theories, especially for those with high-dimensional theory parameter spaces: in Ref.~\cite{Strege:2014ija}, for instance, it was estimated that $\approx 400$ CPU-years would be needed to obtain a statistically convergent scan of a 15-dimensional supersymmetric model.



We demonstrate here that this bottleneck can be bypassed by introducing machine learning (ML) tools that can {\it learn} the mapping between theory and data, and then rapidly and accurately predict signal region efficiencies. 

\header{Gaussian processes}
The number of events $N_i$ in SR $i$ can be written as $N_i = L\sigma \epsilon_i$, where $L$ is the integrated luminosity, $\sigma$ the production cross-section of the relevant process(es), and $\epsilon_i \in [0,1]$ is the SR efficiency (which is in turn the product of the detector efficiency times the acceptance, i.e.~the fraction of events that passes analysis cuts). A {\it classification} ML method was introduced in Ref. \cite{Caron:2016hib} to predict whether or not a given point in the BSM theory parameter space is compatible with LHC data. Here, we are interested in the more general {\it regression} problem of estimating the continuous quantities $\epsilon_i$ given the input BSM parameters $\theoryparam$, i.e. in modeling the relationship $\epsilon_i = f_i(\theoryparam)$. 

We specifically implement here the Gaussian process (GP) regression model~\cite{Rasmussen:2005:GPM:1162254}. Instead of predicting a single value, a GP has the virtue of equipping predictions with consistent uncertainty estimates by means of a full posterior distribution.
The crucial ingredient of GPs is the covariance function, 
which specifies the correlation structure between the function value at different points in the input parameter space. We use here for the covariance function an anisotropic squared exponential kernel \cite{Rasmussen:2005:GPM:1162254}
\ifnotwordcount
\begin{equation}
  k(\theoryparam, \theoryparam') = \sigma_f^2 \exp\left(\sum_j \frac{(\theta_j - \theta'_j)^2}{2l_j^2}\right)
\end{equation}
\fi
where the sum is over the BSM theory parameters.
$\sigma_f$ and $l_j$ are hyperparameters: $\sigma_f^2$ encodes the intrinsic variance of the function we are modeling, and the $l_j$ are characteristic length-scales which determine how quickly the function changes from point to point.
Choosing the optimal values of these hyperparameters to model our function is the \emph{learning} task of GPs, and is done by the standard procedure of evidence maximization~\cite{Rasmussen:2005:GPM:1162254}.

The major limitation of standard GPs is that training scales cubically with the size $n$ of the training data set as it involves computing the inverse of $n\times n$ matrices.
Therefore, in practice, there is a limitation on the amount of training data that can be used.
To eliminate this limitation we make use of distributed GPs (DGPs), specifically the robust Bayesian Committee Machine~\cite{dgp} algorithm, which avoids large matrices by  partitioning the training data into smaller data sets and distributing the computation across independent computing nodes.

\header{Natural supersymmetry}
As a proof-of-concept, we apply this new technique to the natural supersymmetry (SUSY) scenario, in which fine-tuning is low, and the electroweak scale is stabilized by a small subset of light SUSY states (e.g., \cite{Feng:1999mn, Kitano:2006gv, Baer:2011ec, Papucci:2011wy, Baer:2012uy}).
We focus in particular on the minimal natural SUSY scenario of Refs.~\cite{Drees:2015aeo,Kim:2016rsd}, a realistic, yet low-dimensional theory, in which the gluinos, both stops, the left handed sbottom, and the higgsinos all have masses at TeV scale while the remaining states are decoupled.
The six parameters of minimal natural SUSY are:
the supersymmetric Higgs mixing parameter $\mu$,
the gluino mass parameter $M_3$,
the ratio of the two Higgs vacuum expectation values $\tan \beta$,
the third generation SU(2)-doublet squark soft-breaking parameter $m_{Q_3}$,
the third generation SU(2)-singlet soft-breaking parameter $m_{t_R}$,
and the top trilinear soft-breaking term $A_t$.

\header{Data} 
The experimental scenario we consider is the planned high luminosity upgrade of the LHC (HL-LHC)~\cite{lhcplan} with \num{3000} fb$^{-1}$ worth of data collected at \num{14} TeV center-of-mass energy.
We focus on two mutually exclusive SRs defined in Ref.~\cite{atlas-phys-pub-2013-011}, for which the ATLAS collaboration provides background estimates.\footnote{In Ref.~\cite{atlas-phys-pub-2013-011} they vary the cuts as a function of the stop mass.
We use the cuts optimized for \num{1.1} TeV.}
These SRs are optimized for direct production of stops, the most relevant production channel for natural SUSY.
The typical decay channels for the stop are: top or bottom quarks, {\it W}/{\it Z}/Higgs bosons, and the lightest neutralino.
The detector signature is the presence of several jets (including {\it b}-jets), large missing transverse energy, and possibly leptons.
We refer to the ATLAS note for the full definitions of the SRs, and we focus here on the 0-lepton and a 1-lepton SR.

\header{Training and testing}
For training and test data we analyzed \num{18647} samples generated in Ref.~\cite{Kim:2016rsd}, for which SR efficiencies were calculated using \sw{SPheno 3.2.4}~\cite{Porod:2011nf}, \sw{Pythia 8.210}~\cite{Sjostrand:2014zea, Desai:2011su} with default parton distribution function set~\cite{Nadolsky:2008zw}, \sw{NLLFAST 3.1}~\cite{Beenakker:1996ch, Beenakker:1997ut, Kulesza:2008jb, Kulesza:2009kq, Beenakker:2010nq, Beenakker:2011fu}, and \sw{CheckMATE 1.2.1}~\cite{Drees:2013wra, Kim:2015wza,CM2, checkmatewebpage} with \sw{Delphes 3.10}~\cite{deFavereau:2013fsa} and \sw{Fastjet 3.0.6}~\cite{Cacciari:2005hq, Cacciari:2008gp, Cacciari:2011ma}.

We used \num{16647} of these samples to train DGPs for the two SRs, with one single level architecture with an ensemble of \num{256} GP. Training was fast due to the use of the DGP algorithm and took approximately \num{15} minutes on a desktop computer with a 4.0 GHz Intel 4790K processor.
\ifnotwordcount
\begin{figure}
  \centering
  \includegraphics{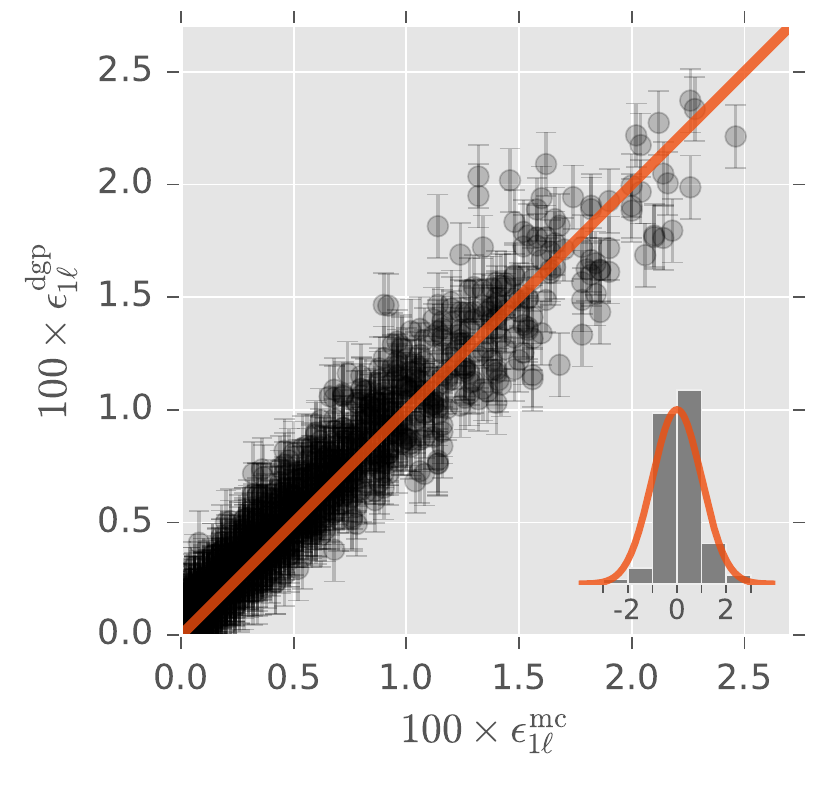}%
\caption{\label{f:test}
The DGP prediction, $\epsilon_{1\ell}^{\mathrm{dgp}}$, versus the MC prediction, $\epsilon_{1\ell}^{\mathrm{mc}}$, for the 1-lepton signal region.
The black circles are \num{2000} test points in the parameter space of natural SUSY.
The errors on $\epsilon_{1\ell}^{\mathrm{dgp}}$ are those predicted by the DGP itself.
The orange line shows the desired behavior $\epsilon_{1\ell}^{\mathrm{dgp}} = \epsilon_{1\ell}^{\mathrm{mc}}$. The insert shows how the distribution of $(\epsilon_{1\ell}^{\mathrm{dgp}} - \epsilon_{1\ell}^{\mathrm{mc}})/\sigma_{1\ell}^{\mathrm{dgp}}$ (gray) compares with the standard normal distribution $N(0,1)$ (orange).}
\end{figure}
\fi
\ifnotwordcount
\begin{figure*}[!t]
  \centering
  \includegraphics{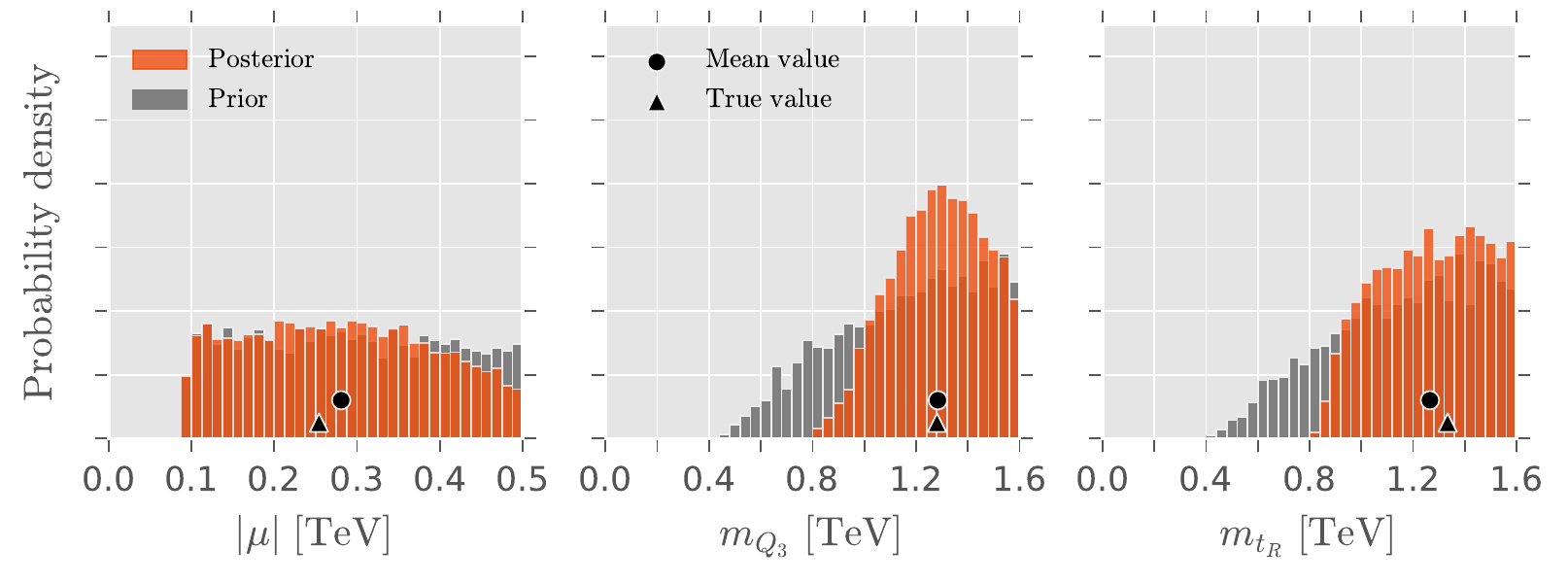}
  \caption{\label{f:oneD}
  Reconstruction of the natural SUSY parameters $\mu$, $m_{Q_3}$, and $m_{t_R}$ using a mock excess generated from the benchmark point. The DGPs was used to calculate the SR efficiencies; no event generation was performed.
  Gray histograms are the marginal priors where we only apply bounds on the Higgs boson mass $\in [121, 129]$ GeV and the chargino mass $> 103.5$ GeV.
  The orange histograms show the marginal posteriors when we also fit to the mock excess.
  We scan over both positive and negative $\mu$ but show only results for $|\mu|$ as they are symmetric.}
\end{figure*}
\fi
We then tested the predictions of the trained DGPs on the remaining \num{2000} points.  In Fig.~\ref{f:test} we show the efficiency predicted by the DGP model in the 1-lepton SR, $\epsilon_{1\ell}^{\mathrm{dgp}}$, versus the values calculated with the full MC calculation, $\epsilon_{1\ell}^{\mathrm{mc}}$, for these \num{2000} test points.
The DGP model accurately predicts the efficiencies, which cluster around the orange line defined by $\epsilon_{1\ell}^{\mathrm{dgp}} = \epsilon_{1\ell}^{\mathrm{mc}}$, with a spread consistent with the DGP error estimate, $\sigma_{1\ell}^{\mathrm{dgp}}$.
We can quantify the agreement by calculating the $\chi^2$; for both the 0-lepton and 1-lepton SRs we get $\chi^2 \approx 1300$, while naively expecting $\chi^2 = 2000 \pm 64$ given the \num{2000} degrees of freedom.
The reason for these low values of $\chi^2$ is that the DGP model slightly overestimates its error.
We visualize this in the insert of Fig.~\ref{f:test} where we see that the distribution of $(\epsilon_{1\ell}^{\mathrm{dgp}} - \epsilon_{1\ell}^{\mathrm{mc}})/\sigma_{1\ell}^{\mathrm{dgp}}$ is more peaked than the standard normal distribution $N(0,1)$. 


\header{Reconstruction}
The DGP model, thus, effectively acts as a surrogate model for the full simulation chain, opening up new opportunities in the interpretation of LHC data. For example, the DGP model can rapidly reconstruct the theory parameters of a BSM model in the case where an excess is observed on top of the Standard Model background.
To demonstrate the feasibility of a full reconstruction procedure, we perform it on an mock data set generated assuming a future excess will be detected.
Our benchmark is the following point in the natural SUSY parameter space: $\mu = 254.6$ GeV, $\tan \beta = 20$, $M_3 = 2000$ GeV, $m_{Q_3} = 1280.5$ GeV, $m_{t_R} = 1333.6$ GeV, $A_t = -2000$ GeV.
The physical masses are \num{1145} and \num{1413} GeV for the stops, \num{1259} GeV for the sbottom, and \num{254} GeV for the neutralino mass.
This benchmark point is not excluded by current searches \cite{Aad:2015pfx,Aaboud:2016lwz,ATLAS-CONF-2016-050,ATLAS-CONF-2016-077}, but should lead to a detectable signal in our signal regions: \num{33.2} ($\sim 3\sigma$ excess over the background) and \num{68.9} ($\sim 4\sigma$) events in the 0-lepton and 1-lepton regions respectively.

As a proof-of-concept, we scan over $\mu$, $m_{Q_3}$, and $m_{t_R}$ -- which are the parameters that govern the masses of the squarks and neutralino involved (the gluino is heavy) -- and fix the other parameters to their benchmark values.
Our priors are uniform over $\mu \in [-0.5, 0.5]$ TeV, $m_{Q_3} \in [0.1, 1.6]$ TeV, and $m_{t_R} \in [0.1, 1.6]$ TeV. We further restrict the Higgs boson mass to the range $121 < m_h < 129$ GeV range\footnote{This range is the $2\sigma$ interval around the central value $m_h = 125.09 \pm 0.24$ GeV \cite{Olive:2016xmw}, with a theoretical error of 2 GeV added in quadrature\cite{Allanach:2004rh}.}, and the chargino mass to be above $103.5$ GeV as per the LEP-2 limit \cite{lep2chargino}. Finally for the ATLAS mock data likelihood construction, we follow the prescription in Appendix A.2 of \cite{Strege:2014ija}. Notice that the 0-lepton and 1-lepton signal regions considered in this work are {\it exclusive}, thus the joint likelihood is the multiplication of the likelihoods for the two signal regions. The uncertainty on the mock signal includes a contribution arising from the uncertainty on the calculation of the cross sections, as well as one arising from the uncertainty on the efficiencies calculated with the DGP model. We neglect correlated systematic uncertainties.
We scan the theory parameter space with a modified version of SuperBayeS\footnote{The core of SuperBayeS is the \sw{MultiNest} scanning algorithm \cite{Feroz:2007kg,Feroz:2008xx} which is interfaced with \sw{SPheno 3.2.4}~\cite{Porod:2011nf} and \sw{NLLFAST 3.1}~\cite{Beenakker:1996ch, Beenakker:1997ut, Kulesza:2008jb, Kulesza:2009kq, Beenakker:2010nq, Beenakker:2011fu} to calculate observables. The likelihood is implemented  in the ROOT framework~\cite{Brun:1997pa} with the RooFit~\cite{Verkerke:2003ir} and RooStats~\cite{Moneta:2010pm} packages.} \cite{deAustri:2006pe, Roszkowski:2007fd, Trotta:2008bp}.

We show in Fig.~\ref{f:oneD} the marginal prior (gray) and the marginal posteriors (orange), i.e. the probability distributions after taking the excess in the mock data into consideration. As one can see, the mock data in the two signal regions have a very limited impact on the determination of $\mu$, but they lead to a measurement on $m_{Q_3}$, and to a more stringent lower limit on $m_{t_R}$.
The posterior slightly disfavors larger values of $|\mu|$ because it determines the neutralino mass, and the missing energy cut requires that the mass difference between the neutralino and the produced squark is large. The parameters $m_{Q_3}$ and $m_{t_R}$, together with the off-diagonal entries of the stop mass matrix, determine the mass of the three light squarks; $\tilde{b}_1$, $\tilde{t}_1$, and $\tilde{t}_2$. The analysis is sensitive to these masses through the production cross-sections, as the total cross-section must be large enough to produce the measured number of events.
The total production cross-section is dominated by two contributions: one that depends on both $m_{Q_3}$ and $m_{t_R}$ (via the mass of $\tilde{t}_1$) and one that depends on $m_{Q_3}$ only (via the mass of $\tilde{b}_1$). The constraint on the total production cross-section therefore translates more directly into a constraint on $m_{Q_3}$, explaining why we are able to reconstruct $m_{Q_3}$ better than $m_{t_R}$.


In reconstructing the benchmark parameters from mock data we used 4 000 live points in \sw{MultiNest}, which required $10^5$ likelihood evaluations before converging. The estimate of the two SR efficiencies with the DGP model took 0.06 seconds per evaluation on a single 4 GHz Intel 4790K core, {\it a factor $\sim 10^4$ faster than the ${\cal O}(10)$ minutes per evaluation required to generate the training set}.
The whole scan took $66$ hours on six CPUs.

\header{Discussion and Conclusions}
In this letter, we have introduced the use of Gaussian processes to accelerate the interpretation of LHC data in the framework of BSM theories. 
Their ability to estimate an error on their predictions makes them ideal for fast and robust approximate calculations. We have specifically demonstrated that the estimate of SR efficiencies can be accelerated by a factor $10^4$, making it possible to rapidly and accurately reconstruct the natural SUSY theory parameters, should an excess in the data be discovered at the HL-LHC. 
The method can be generalized to any BSM theory, and it can be in principle extended to accelerate other time-consuming tasks, such as the calculation of the cross-sections, or of the likelihood itself.

Gaussian processes are currently a very active area of research and it is likely that the method will be further improved and refined. One particularly interesting extension to our implementation is multi-output prediction:
instead of training one (D)GP for each SR separately, one can train a single (D)GP that predicts all the SR efficiencies simultaneously.
The correlation between the SR efficiencies could then be used to make more precise predictions. 
Another direction is to move away from Gaussianity and use Student-t processes~\cite{shah2014student} which might model the underlying noise for the Monte Carlo generators better.
Another intriguing development is Bayesian optimization~\cite{bayesopt}, a form of active learning~\cite{seo2000gaussian, settles.tr90}, which aims to minimize the amount of training data needed, by letting the Gaussian process itself specify where to sample the theory parameter space next in an iterative fashion.

In the traditional approach to global analyses, new samples have to be generated in the parameter space of a BSM theory, and new simulations performed every time new data become available, as the sampling is driven by the likelihood. An important aspect of our new method is that detector simulations need to be performed {\it only once} for each BSM theory, to generate the training sample. Once this is done, the surrogate model can be reused by anyone, and applied to any data set.





\ifnotwordcount
\header{Acknowledgments}
We thank S. Caron and C. McCabe for their useful comments. G.B. (P.I.) and S.L. acknowledge support from the European Research Council through the ERC starting grant {\it WIMPs Kairos}.  M.P.D. acknowledges support from a Google Faculty Research Award.
The work of R. RdA was supported by the Ramón y Cajal program
of the Spanish MINECO and also thanks the support of the grants FPA2014-57816-P and
FPA2013-44773, and the Severo Ochoa MINECO project SEV-2014-0398. He also acknowledge
specially the support of the Spanish MINECO’s Consolider-Ingenio 2010 Programme
under grant MultiDark CSD2009-00064.
JSK wants to thank T. Weber for his help with the efficiency files. The work of JSK was supported by IBS under the project code, IBS-R018-D1 and was partially supported by the MINECO, Spain, under contract FPA2013-44773-P; Consolider-Ingenio CPAN CSD2007-00042 and  the Spanish MINECO Centro de excelencia Severo Ochoa Program under grant SEV-2012-0249.
\bibliography{hepgp}
\fi

\end{document}